\documentclass[a4paper,11pt]{article}
\usepackage{jinstpub}
\usepackage{lineno}
\usepackage{paralist}
\usepackage{wrapfig}
\usepackage{listings}
\lstdefinelanguage{diff}{
  basicstyle=\ttfamily\small,
  morecomment=[f][\color{green!50!black}]{+\ },
  morecomment=[f][\color{red!80!black}]{-\ },
}

\usepackage{lmodern}

\usepackage{mdframed}
\usepackage{todonotes}
\usepackage{placeins}
\usepackage[version=4]{mhchem}
\usepackage{siunitx}

\usepackage{enumitem}
\newlist{inlineenum}{enumerate*}{1}
\setlist[inlineenum]{label=(\roman*), itemjoin={{, }}, itemjoin*={{, and }}}
\setlist[enumerate]{align=left, leftmargin=*}

\usepackage{tikz}
\usetikzlibrary{shapes,positioning}
\usetikzlibrary{arrows,arrows.meta,backgrounds,bending,fit} 

\usepackage{subcaption}
\usepackage[final]{microtype}

\definecolor{_blue}{HTML}{5790fc}
\definecolor{_orange}{HTML}{f89c20}
\definecolor{_red}{HTML}{e42536}
\definecolor{_purple}{HTML}{964a8b}

\title{Accelerating \textsc{Garfield++} with CUDA}
\author[a,1]{T.~Neep,\note{Corresponding author.}}
\author[a,b]{K.~Nikolopoulos,}
\author[a]{M.~Slater}
\affiliation[a]{School of Physics and Astronomy, University of Birmingham, B15 2TT, Birmingham, United Kingdom}
\affiliation[b]{Institute of Experimental Physics, University of Hamburg, 22761, Hamburg, Germany}
\emailAdd{t.j.neep@birmingham.ac.uk}

\abstract{\textsc{Garfield++} is extensively used within the gaseous
  detector community for comprehensive detector simulations,
  supporting the full experimental life cycle from design to operation
  and calibration. The emergence of micro-pattern gaseous detectors
  has necessitated computationally intensive microscopic avalanche
  simulations. The acceleration of one of \textsc{Garfield++}’s most
  demanding algorithms, \textsc{AvalancheMicroscopic}, by porting it to
  graphics processing units using NVIDIA's CUDA framework is
  described. The modifications are integrated into the
  \textsc{Garfield}++ codebase and are accessible to end users with
  only minor adjustments to their existing code. Benchmark results
  demonstrate substantial speed-up, especially for high-gain
  avalanches involving thousands of electrons, thereby enabling more
  efficient and detailed detector simulations.}


\arxivnumber{2509.15377}

\begin{document}
\maketitle
\flushbottom

\section{Introduction}
\label{sec:intro}
\textsc{Garfield}~\cite{Veenhof:1993hz,Veenhof:1998tt}, and its object-oriented C\texttt{++}
reincarnation \textsc{Garfield++}~\cite{garfieldpp},
is the ``de facto standard'' software toolkit for the simulation of
gaseous detectors, and is used widely in the community throughout the
life cycle of experiments.
A challenge in the modelling of gaseous detectors is the multiple
length scales involved. The physical dimensions of the detectors range
from $0.01$ to $1\;\si{\metre}$, while the detector micro-physics
occurs at distances below $10\;\si{\micro\metre}$. For traditional
gaseous detectors, such as drift tubes and multi-wire proportional
counters, the simulation of detector response was based on the use of
electron and ion drift lines and the macroscopic transport
coefficients.

The invention of micro-pattern gaseous
detectors~\cite{Giomataris:1995fq, Bouclier:1996im} made this approach
insufficient. As a result, since 2008 \textsc{Garfield/Garfield++} has been
enhanced with microscopic tracking
algorithms~\cite{Veenhof:2009zza,Schindler:2010los}, implemented
through semi-classical Monte Carlo techniques~\cite{HRSkullerud_1968}
utilising the cross-sections of the atomic processes in the
detector~\cite{Biagi:1999nwa}.
More details on these algorithms and examples of early successes in
describing micro-pattern gaseous detectors are provided in
refs.~\cite{Nikolopoulos:2011zza,Schindler:2012wta}.
Typically, in such use cases, 
the electric
field is obtained numerically, e.g. with the Finite Element Method
or the near exact Boundary Element
Method~\cite{Majumdar:2006jf,Majumdar:2009zzb}. The optimisation of
field map element finding using an Octree structure was key to
suppress the corresponding computational cost during charge
transport~\cite{Bouhali:2018jvs}.

\textsc{Garfield++} currently includes three different methods for
 modelling charge transport:
\begin{inparaenum}[a)]
\item Runge-Kutta-Fehlberg method, which integrates the equation of
  motion of the charged particle and calculates the corresponding
  drift line;
\item AvalancheMC, which at each step integrates the equation of
  motion of the charged particle over a number of microscopic
  collisions or a specific distance, adds diffusion effects through
  Monte Carlo, and continues to the next step; and
\item \textsc{AvalancheMicroscopic}, which models transport at the microscopic
  level, collision-by-collision.
\end{inparaenum}
Of these three methods, \textsc{AvalancheMicroscopic} offers the most detailed
simulation of the avalanche and, consequently, is the most
computationally intensive. This also means that it has the greatest
potential benefit for acceleration and is therefore chosen as the
algorithm to focus on with these investigations.

This article describes how graphics processing unit (GPU) support
using NVIDIA's CUDA framework~\cite{cuda} has been added
to \textsc{Garfield++}, including validation of outputs and
comparisons of the performance when running \textsc{AvalancheMicroscopic} on
GPUs versus central processing units (CPUs).  The driving design
consideration was that the implementation should minimally affect the
existing \textsc{Garfield++} code. Therefore, the approach described here is not entirely
focused on algorithmic speed, but on enabling existing
\textsc{Garfield++} users to benefit from hardware acceleration, while
keeping as close as possible to an established and well-tested
codebase.
Thus, the \textsc{Garfield++} code has been adapted such that a
current user can benefit from GPU capabilities with minimal changes to
their existing \textsc{Garfield++} programs. This is in contrast to
other approaches that have either re-implemented some capabilities of
\textsc{Garfield++} in new packages~\cite{Bouhali:2018jvs} or efforts
to develop completely new simulation codes~\cite{Quemener:2021bzz}.
To achieve this, additions needed to be made such that the various data structures and algorithms required could execute on
the GPU. The implementation and features detailed in this paper
have been publicly available in the \textit{master} branch of the
\textsc{Garfield++} codebase~\cite{garfieldpprepo} since June
2024\footnote{Merge commit hash-key:
c7d1b2fa58f21ecb0fbcadf3be4ff80df3b17430} and in version 2025.1.

This article starts by describing the avalanche process and its
implementation in \textsc{Garfield++} (section~\ref{sec:microscopic}),
focusing on suitability for GPU acceleration. In
section~\ref{sec:library_changes} the approach chosen to incorporate
GPU support into the \textsc{Garfield++} codebase is detailed,
including the challenges addressed, while
section~\ref{sec:user_changes} describes the changes required by a
user to run their code on the GPU. Section~\ref{sec:case_studies}
compares the results and performance of the algorithm running on the
CPU versus the GPU for two case studies: a low-gain single-layer gas
electron multiplier (GEM) and a high-gain three-layer GEM. In
section~\ref{sec:support_and_future}, features not yet available on
the GPU and possible future additions are discussed. Finally, a
summary of the findings is presented in section~\ref{sec:conclusion}.

\section{The Avalanche Process and \textsc{AvalancheMicroscopic} in \textsc{Garfield++}}
\label{sec:microscopic}

Gaseous detectors are used in a wide range of applications. They can
deliver sensitivity to small energy deposits by charge multiplication
through an avalanche process. In this process, an electron travelling
through a strong electric field acquires enough energy between two
collisions with the gas molecules to ionise further, thus being
``multiplied'', resulting in detectable charges. The gain of a detector
is defined as the number of electrons after the avalanche process
divided by the number of initial electrons. Experiments
often aim to achieve the highest possible gains, while still operating
stably in the proportionality regime.

For high-gain detectors, the simulation can become computationally
expensive because each secondary electron created in the avalanche
must be tracked individually until termination conditions are
met. Despite the algorithm treating the electrons independently, the
\textsc{AvalancheMicroscopic} class of \textsc{Garfield++} tracks each particle
\textit{sequentially} on the CPU and, thus, the time to simulate an
avalanche is proportional to the number of electrons it contains. This
is an example of an ``embarrassingly parallel'' problem, well-suited
to execution on a GPU, which offers orders of magnitude more, but less
powerful, processing cores than a CPU.

\begin{figure}
  \centering
  \subcaptionbox{\label{fig:parallel_cartoon:perfect}}{
\begin{tikzpicture}[
    squarednode/.style={rectangle, very thick, minimum size=5mm},
  ]
  \node[squarednode, fill=_blue] (node1_1) {};
  \node[squarednode, fill=_orange] (node1_2) [right=of node1_1]{};
  \node[squarednode, fill=_red] (node1_3) [right=of node1_2]{};
  \node[squarednode, fill=_purple] (node1_4) [right=of node1_3]{};

  \node[squarednode, fill=_blue] (node2_1) [below=of node1_1]{};
  \node[squarednode, fill=_orange] (node2_2) [below=of node1_2]{};
  \node[squarednode, fill=_red] (node2_3) [below=of node1_3]{};
  \node[squarednode, fill=_purple] (node2_4) [below=of node1_4]{};

  \node[squarednode, fill=_blue] (node3_1) [below=of node2_1]{};
  \node[squarednode, fill=_orange] (node3_2) [below=of node2_2]{};
  \node[squarednode, fill=_red] (node3_3) [below=of node2_3]{};
  \node[squarednode, fill=_purple] (node3_4) [below=of node2_4]{};

  \node[squarednode, fill=_blue] (node4_1) [below=of node3_1]{};
  \node[squarednode, fill=_orange] (node4_2) [below=of node3_2]{};
  \node[squarednode, fill=_red] (node4_3) [below=of node3_3]{};
  \node[squarednode, fill=_purple] (node4_4) [below=of node3_4]{};

  \draw[->] (node1_1.south) -- (node2_1.north);
  \draw[->] (node2_1.south) -- (node3_1.north);
  \draw[->] (node3_1.south) -- (node4_1.north);

  \draw[->] (node1_2.south) -- (node2_2.north);
  \draw[->] (node2_2.south) -- (node3_2.north);
  \draw[->] (node3_2.south) -- (node4_2.north);

  \draw[->] (node1_3.south) -- (node2_3.north);
  \draw[->] (node2_3.south) -- (node3_3.north);
  \draw[->] (node3_3.south) -- (node4_3.north);

  \draw[->] (node1_4.south) -- (node2_4.north);
  \draw[->] (node2_4.south) -- (node3_4.north);
  \draw[->] (node3_4.south) -- (node4_4.north);

\end{tikzpicture}
  }
\hspace{6em}
\subcaptionbox{\label{fig:parallel_cartoon:real}}{
  \begin{tikzpicture}[
      squarednode/.style={rectangle, very thick, minimum size=5mm},
    ]
    \node[squarednode, fill=_blue] (node1_1) {};
    \node[squarednode, draw=_orange] (node1_2) [right=of node1_1]{};
    \node[squarednode, draw=_red] (node1_3) [right=of node1_2]{};
    \node[squarednode, draw=_purple] (node1_4) [right=of node1_3]{};

    \node[squarednode, fill=_blue] (node2_1) [below=of node1_1]{};
    \node[squarednode, fill=_orange] (node2_2) [below=of node1_2]{};
    \node[squarednode, draw=_red] (node2_3) [below=of node1_3]{};
    \node[squarednode, draw=_purple] (node2_4) [below=of node1_4]{};

    \node[shape=regular polygon, regular polygon sides=8, very thick, minimum size=5mm, fill=_blue] (node3_1) [below=of node2_1]{};
    \node[squarednode, fill=_orange] (node3_2) [below=of node2_2]{};
    \node[squarednode, fill=_red] (node3_3) [below=of node2_3]{};
    \node[squarednode, draw=_purple] (node3_4) [below=of node2_4]{};

    \node[squarednode, fill=_blue] (node4_1) [below=of node3_1]{};
    \node[squarednode, fill=_orange] (node4_2) [below=of node3_2]{};
    \node[squarednode, fill=_red] (node4_3) [below=of node3_3]{};
    \node[squarednode, fill=_purple] (node4_4) [below=of node3_4]{};

    \draw[->] (node1_1.south) -- (node2_1.north);
    \draw[->] (node2_1.south) -- (node3_1.north);
    \draw[->] (node2_2.south) -- (node3_2.north);
    \draw[->] (node3_2.south) -- (node4_2.north);
    \draw[->] (node3_3.south) -- (node4_3.north);
    \draw[dashed,->] (node1_1.south) -- (node2_2.north);
    \draw[dashed,->] (node2_2.south) -- (node3_3.north);
    \draw[dashed,->] (node3_3.south) -- (node4_4.north);
    \draw[dashed,->] (node3_2.south) -- (node4_1.north);

  \end{tikzpicture}
    }
\caption[GPU thread occupancy]{%
An illustration of GPU thread occupancy in the
(a) 
ideal case and
(b) 
an example \textsc{AvalancheMicroscopic}
case. Rows indicate iteration number and each column represents a GPU thread.
Squares are filled when work is being done, and empty when idle. Solid
arrows represent the continuation of work on a GPU thread and dashed
arrows represent new electrons being created in the avalanche process.
The filled octagon represents the tracking of an electron
terminating.\label{fig:parallel_cartoon}}
\end{figure}
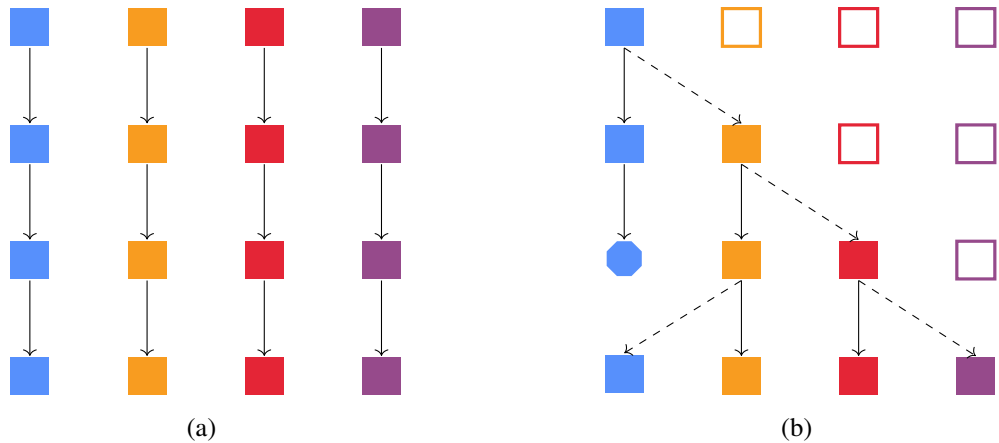

Nevertheless, leveraging GPUs introduces specific challenges that, if
not properly handled, may offset these improvements:
\begin{itemize}
\item The highest speedup is obtained when the GPU cores are fully
  utilized. At the beginning and end of the algorithm there will be
  far fewer electrons to track, thus, many GPU cores will be idle.
  Figure~\ref{fig:parallel_cartoon} shows a schematic comparison of
  the ideal case, where each GPU thread is occupied throughout, with a
  more realistic case for the microscopic avalanche at the avalanche
  growing stage.
\item Both before and after running the algorithm, data needs to be
  transferred to and from the GPU's video RAM. Though fast, this is
  significantly slower than normal memory access. This is because GPU
  memory access prioritizes high throughput for parallel processing,
  tolerating higher latency, while CPU memory access focuses on low
  latency to minimize delays for sequential tasks.
\item Due to the architecture of the GPU cores, highly branched code
  with many conditional statements can cause significant parts of the
  code to be run sequentially instead of in parallel. The
  \textsc{MicroscopicAvalanche} code of \textsc{Garfield++} does
  contain several of these branching paths. However, as the aim of
  this conversion is to minimise material changes to the code,
  alterations were not made to directly reduce this effect.
\end{itemize}

Part of this work is to determine whether the advantages of using GPUs
for simulating the avalanche outweigh these disadvantages, under the
constraint of minimising impact to the original codebase, and if so by
how much.

A rough outline of the \textsc{AvalancheMicroscopic} algorithm is
shown graphically in figure~\ref{fig:avalanche_microscopic_outline}.
The algorithm can be split into two main parts, which are referred to
as \textit{electron transport} and \textit{stack processing}. The
algorithm runs until all electrons under consideration have a non-zero
status code, i.e. they are not to be transported any further, at which
point the avalanche is completed.

The \textit{electron transport}
part of the avalanche can be performed for every ``active'' electron,
i.e. electrons still being transported, in parallel and is, therefore,
where the most benefits are likely to be found from running on the GPU. 
However, these benefits could be cancelled out as a consequence of
the stochastic nature of the algorithm.
Interactions in the code are determined by a combination
of conditionals and random number generation and will consequently be
different during the transport of each electron. This variability
between threads could lead to significant parts of each iteration being
run sequentially on the GPU or large numbers of cores not being
utilised for the whole transport step, both of which could
reduce performance.

After each \textit{electron transport} step has been performed, the
entire stack of electrons is processed. Electrons that are no longer
considered ``active'' are removed from the stack and any newly created
electrons are added. In the CPU code, this \textit{stack processing}
stage can be efficiently performed using STL containers and
algorithms. However, those are not available in CUDA and this kind of
memory manipulation can become computationally expensive on the GPU
if implemented poorly. To get around this, the \textit{thrust}
library~\cite{thrust} was used to perform these stack manipulation
processes in parallel on the GPU. Combining this with a small
modification to the stack processing algorithm to minimise copying of
data structures, resulted in a reduction of the time required for the
\textit{stack processing} stage to a negligible level when run on the
GPU.

\tikzset{
    decision/.style={
        diamond,
        draw,
        text width=10em,
        text badly centered,
        inner sep=0pt,
        aspect=4
    },
    descr/.style={
        fill=white,
        inner sep=2.5pt
    },
    connector/.style={
        -latex,
        font=\scriptsize
    },
    rectangle connector/.style={
        connector,
        to path={(\tikztostart) -- ++(#1,0pt) \tikztonodes |- (\tikztotarget) },
        pos=0.5
    },
    rectangle connector/.default=-2cm,
    straight connector/.style={
        connector,
        to path=--(\tikztotarget) \tikztonodes
    }
}

\begin{figure}
  \centering
  \begin{tikzpicture}[
    boxnode/.style={rectangle, draw=black, thick, minimum height=1cm, node distance=1cm},
  ]

    \node[boxnode] (input) {Initial input electron position(s)};
    \node[boxnode] (next_electron) [below=of input] {Next electron};
    \node[decision] (check_pos) [below=of next_electron] {Position is valid?};
    \node[boxnode] (find_efield) [below=of check_pos] {Find $\vec{E}$ field and add random time step};
    \node[decision] (collision) [below=of find_efield] {Collision?};
    \node[decision] (process) [below=of collision] {Collision is ionisation or attatchment?};
    \node[boxnode] (store_updated) [below=of process] {Store updated electron info};
    \node[decision] (all_done) [below=of store_updated] {All electrons processed?};
    \node[boxnode] (remove_dead) [below=of all_done] {Remove inactive and add new electrons};

    \draw[>=latex,->] (input.south) -- (next_electron.north);
    \draw[>=latex,->] (next_electron.south) -- (check_pos.north);
    \draw[>=latex,->] (check_pos.south) to node[descr] {Yes} (find_efield.north);
    \draw[>=latex,->] (collision.south) to node[descr] {Yes} (process.north);
    \draw[>=latex,->] (find_efield.south) -- (collision.north);
    \draw[>=latex,->] (process.south) to node[descr] {Yes} (store_updated.north);
    \draw[>=latex,->] (store_updated.south) -- (all_done.north);
    \draw[>=latex,->] (all_done.south) to node[descr] {Yes} (remove_dead.north);


    \draw [rectangle connector=4cm] (all_done.east) to node[descr] {No} (next_electron.east);
    \draw [rectangle connector=-2cm] (check_pos.west) to node[descr] {No} (store_updated.west);
    \draw [rectangle connector=2cm] (collision.east) to node[descr] {No} (find_efield.east);
    \draw [rectangle connector=2.5cm] (process.east) to node[descr] {No} (find_efield.east);

    \scoped[on background layer]
      \node (bkg_el) [dashed, draw=gray, very thick, inner xsep=4cm, inner ysep=2mm,
                      fit=(next_electron) (check_pos) (process) (store_updated) (all_done)] {};
    \scoped[on background layer]
      \node (bkg_el) [dotted, draw=gray, very thick, inner xsep=4cm, inner ysep=2mm, fit=(remove_dead)] {};

  \end{tikzpicture}
  \caption{A simplified outline of the \textsc{AvalancheMicroscopic}
    algorithm. The \textit{electron transport} and \textit{stack
      processing} stages of the algorithm are represented by the
    dashed and dotted boxes, respectively. The entire algorithm is
    repeated until there are no active electrons remaining.\label{fig:avalanche_microscopic_outline}}
\end{figure}
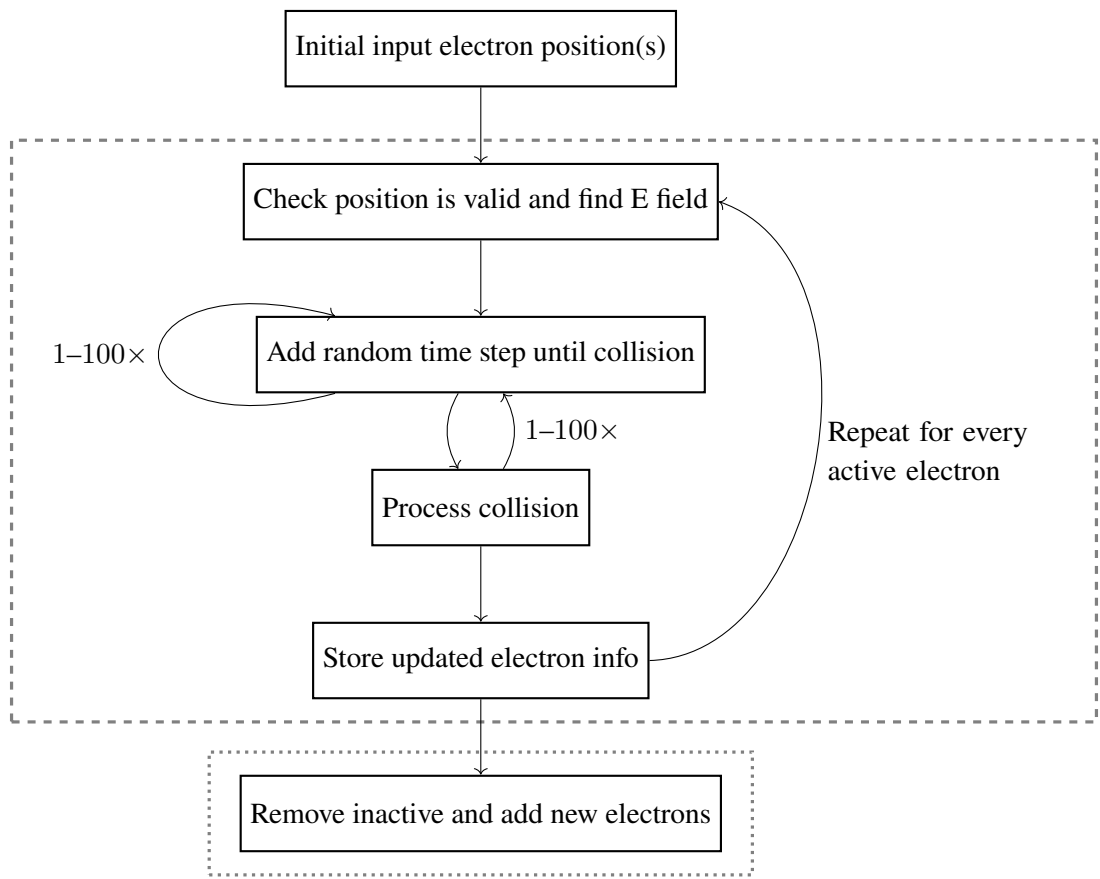

\section{Implementing the changes}

\subsection{Library changes}
\label{sec:library_changes}

Though the core algorithmic code for the simulation can run unchanged
on the GPU, managing both the data structures and code hierarchy proved
more challenging due to the original code's extensive
use of both dynamic polymorphism and STL containers.

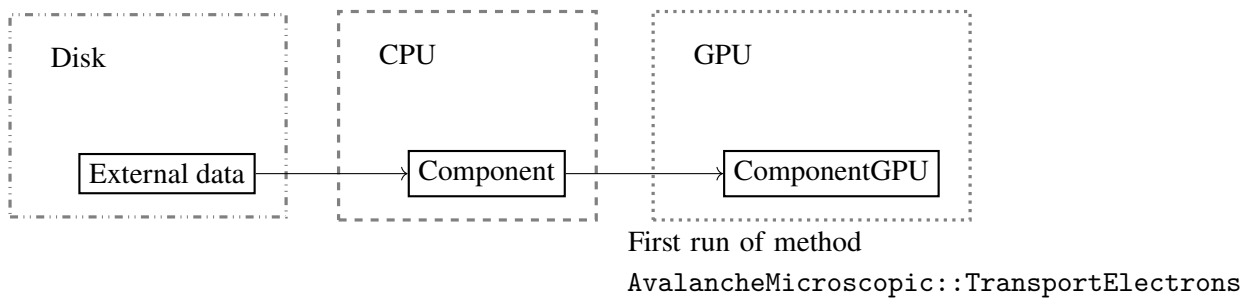
\begin{figure}
  \centering
  \begin{tikzpicture}[
    squarednode/.style={rectangle, draw, thick, minimum size=5mm},
  ]
    \node[squarednode] (cpu) {Component};
    \node (cpu_label) [above=of cpu.north west] {CPU};
    \node[squarednode] (gpu) [left=-7cm of cpu] {ComponentGPU};
    \node (gpu_label) [above=of gpu.north west] {GPU};

    \node[squarednode] (cpu_data) [left=2cm of cpu] {External data};
    \node (cpu_data_label) [above=of cpu_data.north west] {Disk};

    \draw[->] (cpu.east) -- (gpu.west) node [below=1.2 cm of cpu.south west, pos=0.33, anchor=west, text width=8cm]
         {First run of method\\\texttt{AvalancheMicroscopic::TransportElectrons}};
    \draw[->] (cpu_data.east) -- (cpu.west);

    \scoped[on background layer]
    \node (bkg_disk) [dashdotdotted, draw=gray, very thick, inner xsep=4mm, inner ysep=3mm, fit=(cpu_data) (cpu_data_label)] {};

    \scoped[on background layer]
    \node (bkg_cpu) [dashed, draw=gray, very thick, inner xsep=4mm, inner ysep=3mm, fit=(cpu) (cpu_label)] {};

    \scoped[on background layer]
    \node (bkg_gpu) [dotted, draw=gray, very thick, inner xsep=4mm, inner ysep=3mm, fit=(gpu) (gpu_label)] {};
  \end{tikzpicture}
  \caption{An example of the flow of data from the CPU to the
    GPU. Data is loaded from disk into the \texttt{Component} class on
    CPU accessible memory. The data is transfered to the
    \texttt{ComponentGPU} class when the \texttt{TransportElectrons}
    method of \textsc{AvalancheMicroscopic} is first run.  At this
    point the sizes of the data structures are known.\label{fig:gpu_class_data_flow}}
\end{figure}

As mentioned above,
traditional STL containers are not available in CUDA, but whereas for the
stack processing the containers were required to be dynamic, the ones used
in the main algorithmic code were essentially static. Special ``GPU'' versions
of these classes could therefore be written that contained all the required data but in
standard C-style arrays and structures. When the
classes are created, their data is initialised using the standard CPU
version of the class, with as much preprocessing as possible performed
on the CPU before transferring data to the GPU. An illustration of
such a class and the flow of data is shown in
figure~\ref{fig:gpu_class_data_flow}. The total volume of data
transferred from these classes depends on the complexity of the electric field
maps being used, but ranges from around 100~MB to 2~GB in the example
discussed in section~\ref{sec:case_studies}.

The polymorphism used throughout the  \textsc{Garfield++} codebase also proved to be problematic
because, though CUDA can build C\texttt{++} classes, the use of virtual methods can be
very slow on the GPU due to the expense of the VTable lookups. To avoid this problem
but still maintain the ability to call different functions depending on the class,
enumerations of the class type were added that could be easily and efficiently checked
in a flat class hierarchy.

As the work undertaken so far has focused on ease-of-use for the end
user, rather than performance, the majority of code in the CPU and GPU
versions of these classes is kept exactly the same, with preprocessor
macros used to handle differences between valid CUDA and C\texttt{++}
code. This keeps code repetition to a minimum, while simultaneously
clearly marking which methods of each class have been edited to add
GPU support. The majority of changes, as already mentioned, are
replacing the use of C\texttt{++} standard library containers
e.g. \texttt{std::vector} and \texttt{std::array}, with CUDA
compatible or C-style versions. The CUDA code has been compiled and tested with CUDA version 12.1.1.

\subsection{User changes}
\label{sec:user_changes}

The changes required by the user are kept to a minimum. The user needs
to add just a single line when setting up their \textsc{AvalancheMicroscopic}
class in order to run the avalanche on the GPU. The first argument to
\texttt{SetRunModeOptions} controls whether the avalanche simulation
is performed on the GPU (\texttt{MPRunMode::GPUExclusive}) or on the
CPU (\texttt{MPRunMode::Normal}), which is the default. The second
argument, set to \texttt{0} in the example below, selects which GPU
the calculation is run on; useful in case a system has multiple GPUs.
\begin{mdframed}
\begin{lstlisting}[language=diff]
  AvalancheMicroscopic aval;
  aval.SetSensor(&sensor);
+ aval.SetRunModeOptions(MPRunMode::GPUExclusive, 0);
\end{lstlisting}
\end{mdframed}

After the avalanche has been performed, one can then retrieve the electron endpoints with
\begin{mdframed}
\begin{lstlisting}[language=diff]
  unsigned int endpoints =
-     aval.GetNumberOfElectronEndpoints();
+     aval.GetNumberOfElectronEndpointsGPU();
  double xe1, ye1, ze1, te1, e1;
  double xe2, ye2, ze2, te2, e2;
  int status;
  for (unsigned int i=0; i<endpoints; ++i) {
-     aval.GetElectronEndpoint(
+     aval.GetElectronEndpointGPU(
          i,
          xe1, ye1, ze1, te1, e1,
          xe2, ye2, ze2, te2, e2,
          status
      );
  }
\end{lstlisting}
\end{mdframed}
Induced signals can be accessed in the normal way.

As discussed in section~\ref{sec:case_studies}, for large avalanches sizes, typically exceeding more than $10^4$ electrons, the GPU outperforms the CPU. However, for small avalanche sizes, as a result of the overhead for copying electron data to GPU memory, the CPU could still perform better than the GPU. For this reason,  the option \texttt{MPRunMode::GPUWhenAppropriate} has been introduced to 
\texttt{SetRunModeOptions}, which enables the automatic switching between the CPU and GPU when the avalanche increases beyond a user defined cross-over point.

\subsection{Validation}
\label{sec:validation}

It is critical that the changes to the codebase have no or minimal
impact on the numerical output of the algorithm. To perform this check
requires fixing all input parameters between CPU and GPU. The primary factor
is the different Psuedo-Random Numbers used by Garfield (the ROOT random number generator, TRandom3) and the GPU (CUDA library cuRAND). The specifics of the
random number generator are not relevant to the event generation and consequently for
validation,  a pre-generation strategy is used to ensure identical random
numbers during event simulation. A large number of
random number sets totalling 14.5GB are pre-generated and transferred
to the GPU. During event generation, both CPU and GPU code can then
draw from the same random numbers, with the specific set determined by
electron ID in order to avoid differences as a consequence of the
parallel threads on the GPU.

Using these sets of random numbers and the same initial conditions, the same single large avalanche event 
could be generated on both the CPU and GPU simultaneously and any differences checked to a very high precision.
The event used for this validation has an avalanche size of $1.2\cdot 10^5$ 
and 160 total iterations of the transport loop.
No differences are seen between the CPU and GPU generations until just after the peak of the avalanche at which point a slight
deviation of 0.15\% in size develops.

Looking at the errors in position for specific electrons as 
they are tracked through the avalanche shows differences at the $10^{-8}$ level between the CPU and GPU. This is accounted for due
to differences in the order of floating point operations generated by the different compilers and architectures. However, this very
small difference grows through the generation until it creates the more noticeable difference seen in the avalanche size.
This is an irreducible error but the validation gives confidence that the algorithm is performing on the GPU in the
same way and using the same calculations as the original CPU code.

\section{Case studies}
\label{sec:case_studies}

In order to test the performance of the GPU implementation two case
studies are investigated. The first test case uses a low-gain
single-layer GEM, while the second test case uses a high-gain
triple-layer GEM. These test cases were chosen as the field maps
required are already part of the  \textsc{Garfield++} codebase and are used in
existing examples (in
\href{https://gitlab.cern.ch/garfield/garfieldpp/-/tree/master/Examples/Gem}{\texttt{Examples/GEM}}
and
\href{https://gitlab.cern.ch/garfield/garfieldpp/-/tree/master/Examples/Ansys123}{\texttt{Examples/Ansys123}},
respectively). Both the single- and triple-GEM field maps are created
using ANSYS~\cite{ANSYS} and the same gas settings used in the
existing examples are used, with the exception of Penning transfer
which is disabled as it is not yet supported on the GPU.

In each case study, to test performance of the GPU implementation with
respect to the CPU implementation, the time taken to process the
entire avalanche is evaluated for several different CPU and GPU
models. In the case of the GPU this includes the time taken
transferring the initial electron positions to the GPU, but not the
overhead of transferring the field map data (which only has to be done
once). For clarity, only the best performing CPU of those tested is
shown in figures.

\begin{figure}
    \centering
    \subcaptionbox{\label{fig:endpoints}}{\includegraphics[height=0.45\textwidth]{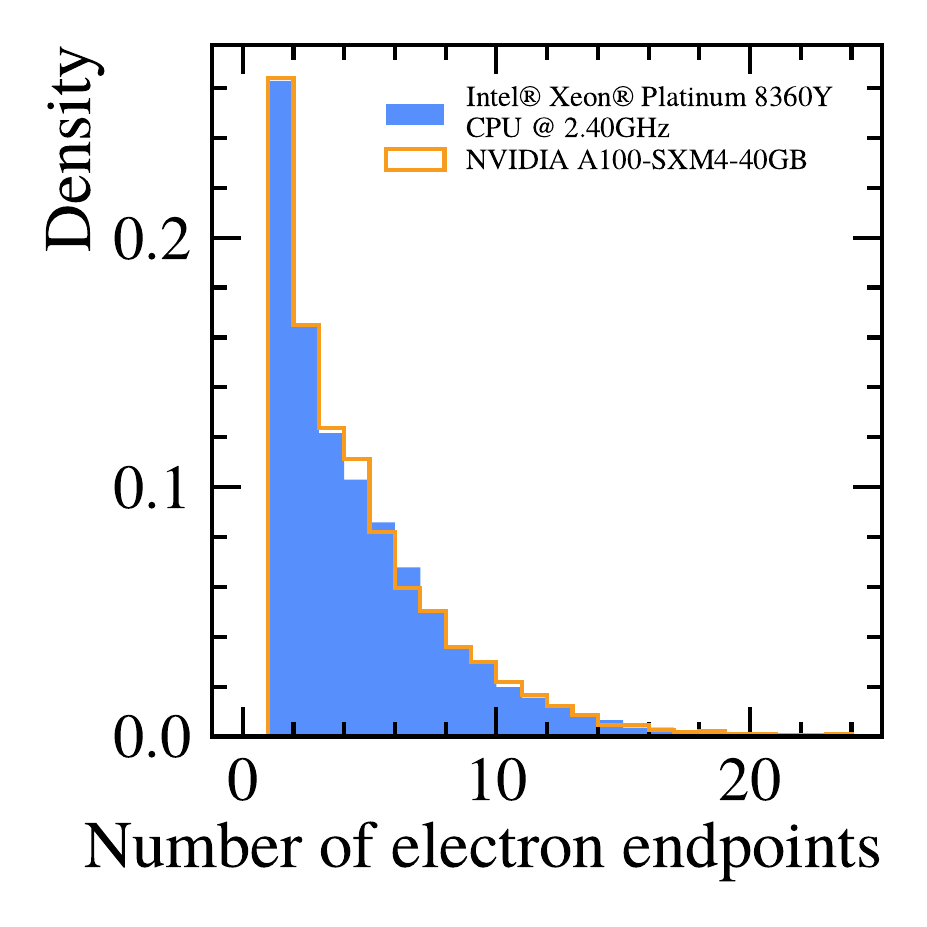}}\hfill
     \subcaptionbox{\label{fig:cpu_v_gpu_time}}{\includegraphics[height=0.45\textwidth]{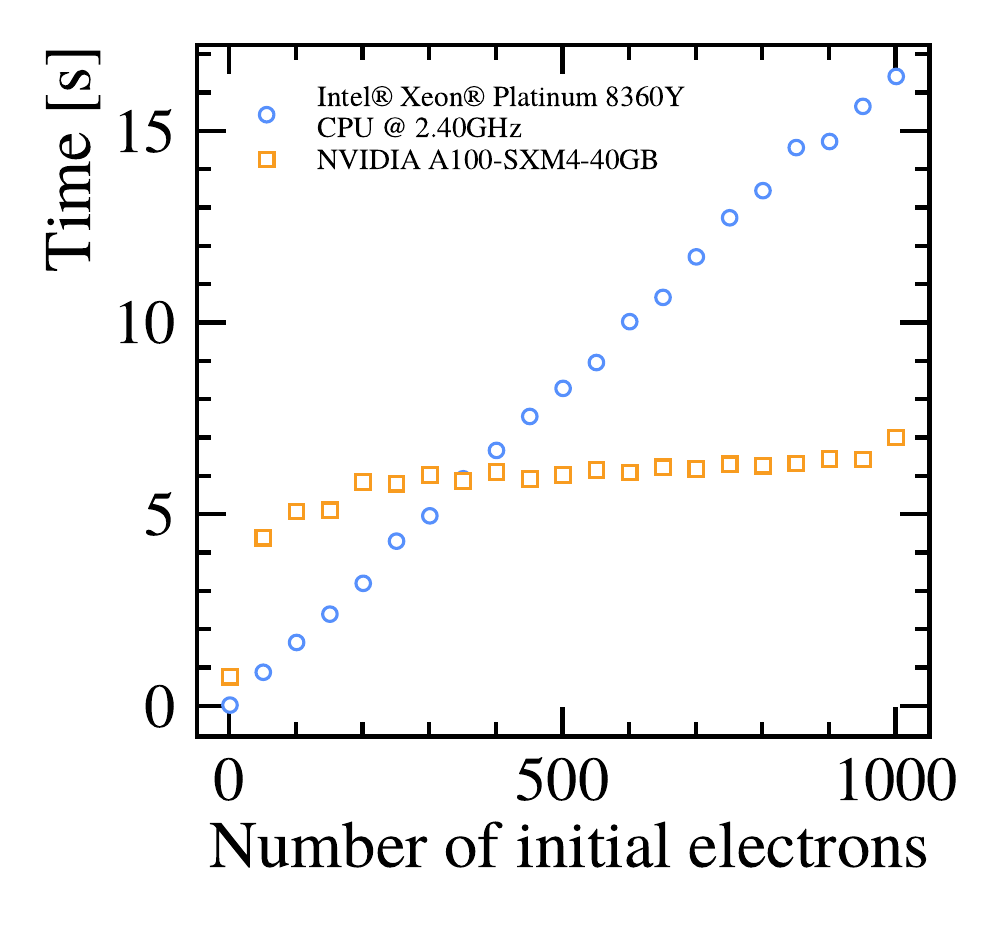}}
     \caption{(a) 
       Number of electron endpoints for 5,000 single electron events.
       (b) 
       Time to run the single GEM example for different numbers of starting electrons (averaged over several runs). The small scatter in points is due to fluctuations in the avalanche process.
    }
\end{figure}

\subsection{A single GEM}
\label{sec:single_gem}
To compare the compatibility of the CPU and GPU versions of the
avalanche, single electron events are simulated with the electron
starting at a random position 0.2 mm above the GEM plane. A histogram
showing the number of endpoints for each event is shown in
figure~\ref{fig:endpoints}. The number of
endpoints produced in the CPU and the GPU agree, with the small differences observed being due to the different random number generators used.

To evaluate the performance of the GPU with respect to the CPU, the time to run the avalanche process as a function of the number of initial electrons is measured. The results are shown in figure~\ref{fig:cpu_v_gpu_time}. It can be seen that the time to run the avalanche on the CPU increases linearly with the initial number of electrons. In the case of the GPU there is a ``turn on'', due to copying electron data to GPU memory when the avalanche begins but after that the time taken increases linearly but slower than the
CPU case with increasing number of electrons. The cross-over point when comparing the CPU to the NVIDIA A100 GPU is at approximately $350$ initial electrons. 

\begin{wrapfigure}[14]{r}{0.5\linewidth}
  \centering
  \includegraphics[width=0.99\linewidth]{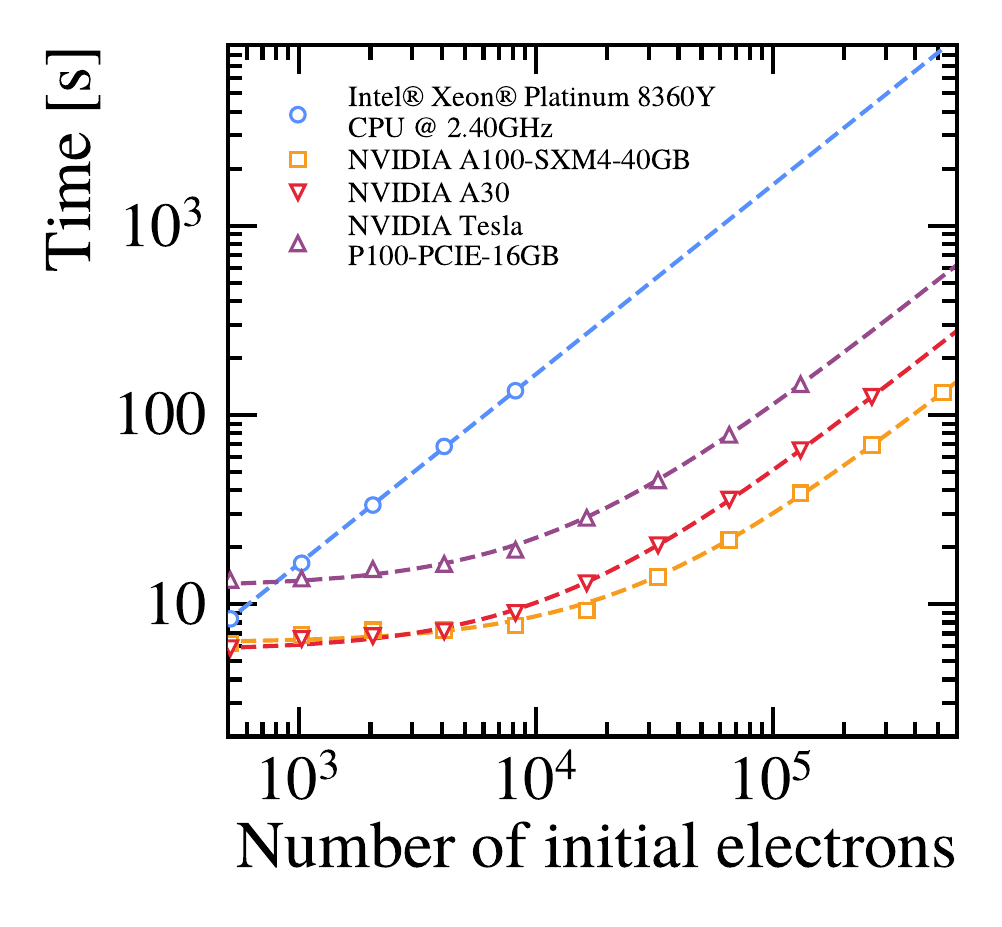}
  \vspace{-2.5em}
    \caption{%
    \label{fig:time_log}
    Time to run the single GEM example. 
    }
\end{wrapfigure}

The results from additional GPU models are shown in
figure~\ref{fig:time_log}, including results with a larger number of
initial electrons. The time taken to process the avalanche increases
linearly with the number of intial electrons for all the models
tested. For a large number of initial electrons the NVIDIA A100 GPU
processes the avalanche nearly 70 times faster than the CPU.  Thanks
to the larger number of initial electrons, this example is closer to
the ideal case of parallelism on the GPU, shown in
figure~\ref{fig:parallel_cartoon:perfect}, than the example with fewer
initial electrons and higher gain, which is discussed next.

\subsection{A triple GEM}
\label{sec:triple_gem}
The second test case considered is that of a triple GEM, where three
GEMs are layered with the goal of increasing amplification. In this
example the average gain is approximately $10^{4}$ for a pressure of 1~bar, 
substantially higher than that of the single GEM example. This example starts with a
single electron. The example is modified from the example included
with  \textsc{Garfield++} to remove an avalanche size limit so that the
avalanche calculation does not terminate early. The figure of merit is
the number of endpoints in the avalanche versus the time taken. In
order to test even higher avalanche sizes the example is further
modified to simulate lower gas pressures, leading to higher average
gains. The high-gain nature of this example is more similar to
real-world use cases of  \textsc{Garfield++} than the single GEM example, and
also more similar to figure~\ref{fig:parallel_cartoon:real}.

\begin{wrapfigure}[15]{r}{0.5\linewidth}
  \centering
    \includegraphics[width=0.99\linewidth]{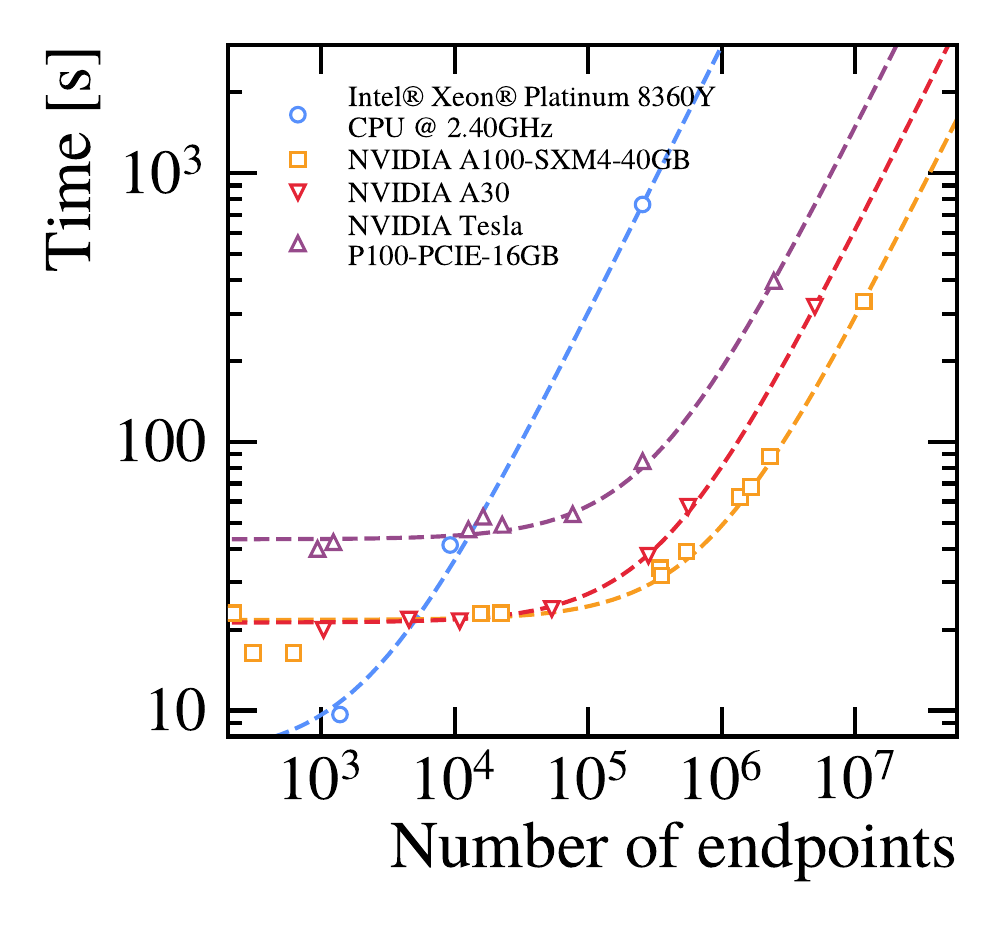}
    \vspace{-2.5em}
\caption{%
    \label{fig:triple_gem_results}
    Time taken for different final avalanche sizes in the triple
    GEM example. 
    }
\end{wrapfigure}

Figure~\ref{fig:triple_gem_results} shows the results. 
The time taken to perform the calculation is proportional to the final size of the avalanche and scales linearly, with the dashed lines showing linear fits to the measurements. As with the example of section~\ref{sec:single_gem} it can be seen the GPU models tested out-perform the CPU once the avalanche size becomes large, approximately $10^4$. At gains of approximately $10^6 \left(10^7\right)$ the avalanche simulation runs approximately $60 \left(100\right)$ times faster on the A100 GPU than on the CPU.

\clearpage

\section{Future directions}
\label{sec:support_and_future}

Currently, running \textsc{AvalancheMicroscopic} on the GPU is available to users with
certain caveats.  The method is not yet available for use in
semiconductors, or in gaseous detectors with a magnetic field,
although there does not appear to be any technical reasons that prevent these cases being
implemented on the GPU.

Beyond adding support for additional features on the GPU, there are a
number of avenues that can be investigated for improvements. GPU
efficiency could be improved by revisiting the algorithms and doing more extensive
changes to reduce branching code
paths. 

Due to the use of the CUDA language, the GPU accelerated algorithms of
 \textsc{Garfield++} require access to an NVIDIA GPU, precluding the use of GPU
models from other manufacturers. As the structural changes required to
offload computation to a GPU have now been made, converting the code
to run on other GPU models should be possible with the use of a
library such as Kokkos~\cite{CarterEdwards20143202,9485033}. However,
particular care would be required due to the use of the CUDA-specific
\textit{thrust} library in the stack processing code.

\section{Conclusion}
\label{sec:conclusion}
Support for running the microscopic avalanche algorithm of  \textsc{Garfield++}
on a GPU has been added. The implementation
in the CUDA programming language allows NVIDIA GPUs to be
exploited, with minimal adjustments to existing user code. The compatibility of the CPU and GPU versions of the
algorithm has been tested and the results are found to agree to high
precision. The parallel nature of processing on the GPU means that large speeds up are possible for  large avalanches, with improvements by factors up to $100$ being observed in the presented
case studies.

This work demonstrates the feasibility of adding GPU support to the
 \textsc{Garfield++} codebase. While not being feature complete yet, it implements
one of the most computationally intensive parts of gaseous detector simulation. The features detailed in this paper have been publicly available in the \textit{master} branch of the  \textsc{Garfield++} codebase since June 2024.

\acknowledgments
The authors gratefully acknowledge Rob Veenhof for his pioneering
development and stewardship of the \textsc{Garfield} simulation program, which
has profoundly influenced research in gaseous detector physics. His
dedication and vision have inspired our work and that of many
colleagues across the scientific community.
The \textsc{Garfield++} maintainers, particularly Heinrich Schindler,
are acknowledged for accepting the developments presented in this
article into \textsc{Garfield++} and for reviewing the relevant merge
requests.

We were saddened to learn of the passing of Stephen Biagi during the final stages of preparing this manuscript. His pioneering work in developing the Magboltz program, a cornerstone tool for calculating electron transport in gases, has had a lasting impact on detector physics. We gratefully acknowledge his profound contributions, which continue to enable advances across our field.

\paragraph{Funding information}
This project has received funding from the European Union’s Horizon
2020 research and innovation programme under the Marie
Skłodowska-Curie grant agreement No 101026519 (GaGARin).
This work was supported in part by the University of Birmingham
Particle Physics consolidated grant No. ST/W000652/1.
KN acknowledges support by the Deutsche Forschungsgemeinschaft (DFG,
German Research Foundation) under Germany’s Excellence Strategy — EXC
2121 “Quantum Universe” — 390833306.
The computations with the "A100-SXM4" GPUs were performed using the
Baskerville Tier 2 HPC service
(https://www.baskerville.ac.uk/). Baskerville was funded by the EPSRC
and UKRI through the World Class Labs scheme (EP/T022221/1) and the
Digital Research Infrastructure programme (EP/W032244/1) and is
operated by Advanced Research Computing at the University of
Birmingham.

\bibliographystyle{JHEP}
\bibliography{bibliography.bib}
\end{document}